\documentclass[5p,times,final,a4paper]{elsarticle}
\usepackage[utf8]{inputenc}
\usepackage{amssymb}
\usepackage{latexsym}
\usepackage{amsmath}

\usepackage[switch]{lineno}% Enable numbering of text and display math, enabled below with \linenumbers

\usepackage{booktabs}
\usepackage{xcolor,soul,ulem}
\usepackage[hidelinks]{hyperref}
\hypersetup{
    colorlinks,
    linkcolor={red!50!black},
    citecolor={blue!50!black},
    urlcolor={black}
}

\definecolor{revAcolor}{HTML}{000000}
\newcommand{\revA}[1]{{#1}}

\definecolor{revBcolor}{HTML}{000000}
\newcommand{\revB}[1]{{#1}}

\journal{Journal of Quantitative Spectroscopy and Radiative Transfer}

\begin{document}
\begin{frontmatter}
\title{%
Optimization of the Femtosecond Laser Impulse for Excitation and the Spin-Orbit Mediated Dissociation in the NaRb Dimer
}
\author[gda1,gda2]{J. Kozicki}
\address[gda1]{Faculty of Applied Physics and Mathematics, Gdańsk University of Technology, 80-233 Gdańsk, Poland}
\address[gda2]{Advanced Materials Center, Gdańsk University of Technology, 80-233 Gdańsk, Poland}
%\ead{jan.kozicki@pg.edu.pl}% ORCID: 0000-0002-8427-7263
\author[gda1,gda3]{P. Jasik\corref{cor1}}
\ead{patryk.jasik@pg.edu.pl}%, ORCID: 0000-0002-6601-0506
\address[gda3]{BioTechMed Center, Gdańsk University of Technology, 80-233 Gdańsk, Poland}
\cortext[cor1]{Corresponding author at Faculty of Applied Physics and Mathematics, Gdańsk University of Technology, 80-233 Gdańsk, Poland.}
\author[gda1,gda2]{T. Kilich}
%\ead{tymon.kilich@pg.edu.pl, ORCID: 0000-0001-6831-694X}
\author[gda1,gda2]{J. E. Sienkiewicz}
%\ead{jozef.sienkiewicz@pg.edu.pl, ORCID: 0000-0002-1149-3846}
%

\date{\today}

%\received{\dots}
%\finalform{\dots}
%\accepted{\dots}
%\availableonline{\dots}
%\communicated{\dots}

\begin{abstract}
\noindent
\begin{minipage}[b]{133.5mm}
We study the
dynamics of multiple coupled states under the influence of an arbitrary
time-dependent external field to investigate the femtosecond laser-driven
excitation and the spin-orbit mediated dissociation in the NaRb dimer.
In this process, the dimer is excited from the ground triplet state
$1^3\Sigma^+$ to the $1^3\Pi$ state using the femtosecond laser impulse and the
spin-orbit coupling between the $1^3\Pi$ and $2^1\Sigma^+$ states results in
the singlet-triplet transition.
The laser impulse parameters are optimised to obtain maximum yield in electronic
states correlating with the first excited atomic asymptote. We observe the
detailed population statistics and power-law decay of these states. Finally, the
analysis of the population oscillations allows us to determine the optimal time
delay for dumping the molecule to its absolute ground state.
\end{minipage}%
\hbox{\vbox{\noindent\includegraphics[height=4cm,width=5cm]{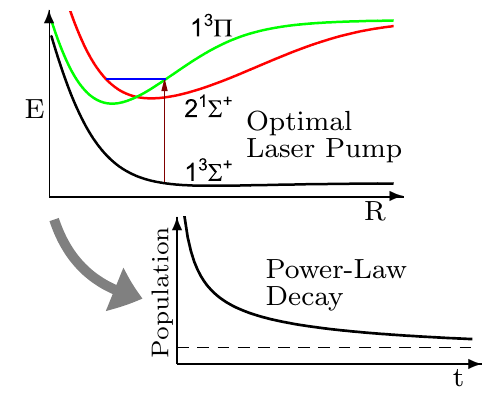}\\
\vbox{\hspace{10mm}Graphical Abstract}}}\\[3mm]
\end{abstract}

%\begin{keyword}
%quantum dynamics \sep field-molecule interaction \sep photoassociation \sep power-law decay \sep diatomic molecule \sep spin-orbit coupling \sep transition dipole moment
%\end{keyword}
\end{frontmatter}

%%%%%%%%%%%%%%%%%%%%%%%% ENABLE LINE NUMBERS
%\linenumbers\relax
\section{\label{sec:level1} Introduction}

Quantum dynamics is a growing discipline at the interface of chemistry, physics
and materials science~\cite{Zewail2000}. It allows studying the behaviour of
objects in a way, that emphasizes the quantum nature of their evolution in time.
Quantum dynamics simulations are an indispensable tool for investigating
processes such as chemical reactions~\cite{Zhao2021_Nature}, field-molecule
interactions~\cite{Yuan2017} and quantum
computing~\cite{Ollitrault2021,Jahangiri2021}. The particular emphasis
is placed on investigating the photoinduced dynamics of breaking (dissociation)
\cite{Wei2017, BAI2020, Vinklarek2021, Chang2021, Zhao2021, Murillo2021} and
creating (association) \cite{Koch2012, Ulmanis2012, Carini2015, Tomza2019, WANG2020, Weyland2021} of the chemical bonds in the molecular systems.
Photoassociation processes play a key role in the field of cold and ultracold
physics and chemistry, allowing for the formation of molecules in the deeply
bound ground states and investigation of their unique quantum properties
\cite{Borsalino2014, Borsalino2016, Balakrishnan2016, Devolder2021, Leung2021}.
On the other side, controlled photodissociation reactions enable for creation of
atomic and molecular fragments in specific quantum states and use them for
researching the selected properties of matter \cite{Alekseyev2000, Kokkonen2014, 
McDonald2016, Wei2017, Toth2020}. While the dissociation through the barrier
of the potential (tunneling) is well described by an exponential decay
\cite{Tannor2007, Schmidt2015, Jasik2018a}, the dissociation of the system of
coupled excited electronic states should be rather described by other forms of
decays. Several different models for these kinds of decays were investigated,
such as product decay~\cite{Kong2020}, dephasing~\cite{Gruebele1998}, and the
power-law decay~\cite{Mizuno2018, Utterback2019, Dong2015}.

Our study aims to show the time-dependent descriptions of the excitation and
dissociation processes in the polar alkali diatomic molecules and the quantum
properties of these reactions in their characteristic time regimes. We propose
the femtosecond laser impulse parameters' optimization procedure providing the
maximization of the population in the coupled complex of excited electronic
states. Dynamics investigations of this system of molecular states allow us, in
the case of an association, to establish the optimal time delays for the
femtosecond laser impulses enabling the formation of molecules in the
deeply bound ground state, as well as present the generally modified power-law
decay allowing for the proper description of the dissociation process.

Our considerations are based on an exemplary polar alkaline dimer, which is
the NaRb molecule. The sodium rubidium molecular system has been studied both
experimentally \cite{Tamanis2002, Docenko2004, JARMOLA2005, Pashov2006, Docenko2007, Guo2022}
and theoretically \cite{Korek2000, Chaieb2014, Wiatr2015, Wiatr2018}. It is
worth underlining that in a recent experiment~\cite{Guo2016, Guo2017} Guo et al.
reported successful production of an ultracold sample of {${}^{23}$Na${}^{87}$}
Rb molecules. They used the Feshbach resonance formed by magnetoassociation and
populated the lower vibrational levels of the ground state by high-efficiency
two-photon Raman processes.

In recent years, in the field of quantum dynamics several numerical methods and their improvements have been
introduced, including higher-order split operator methods, the expansion of the
time evolution operator to Chebyshev
polynomials~\cite{VanDijk2022,Soley2022,Kosloff1984,Kosloff1997,TalEzer2012}, or
the \textit{semi-global} method~\cite{Schaefer2017}. The effective Hamiltonian
for describing nuclear kinetics in coupled multiple adiabatic or diabatic
electronic states plays a key role in the investigated system. Therefore, in the
present study, we apply our newly-developed code~\cite{Kozicki2023} to simulate the impulse-driven
dynamics of photoassociation of the NaRb dimer that takes the spin-orbit
coupling, which varies with the distance between the nuclei, into consideration.
The dynamics of multiple coupled electronic states is determined by solutions to
the coupled time-dependent Schrödinger equations which also includes an
external, classical time-dependent electric field.

In our study, the NaRb dimer is excited from the ground triplet state
$1^3\Sigma^+$ to the $1^3\Pi$ state using the femtosecond impulse and the
spin-orbit coupling between the $1^3\Pi$ and $2^1\Sigma^+$ states results in the
singlet-triplet transition. In the present analysis, 
we use the lowest electronic states of the NaRb molecule with a rotational quantum 
state fixed at $J=0$, appropriate transition dipole moment functions (TDMF), and 
spin-orbit coupling (SOC) matrix elements. \revA{Although accurate experimental 
potential energy curves and SOCs obtained using deperturbation approach are available in literature, e.g. \cite{Docenko2007}, we use our theoretically derived Born-Oppenheimer potential energy curves (PEC) which were presented in our previous work~\cite{Wiatr2015}, 
in order to keep consistency of our calculations (the same set of effective core potentials, core polarization potentials, and basis sets for Na and Rb atoms, as well as the same active space in the configuration interaction method is used for PEC, SOC and TDMF).} The impulse is modeled via the classical,
time-dependent electric field that interacts with the electronic states via the
dipole moments. \revA{Since our model includes two excited electronic states, excitations to higher states, including two- and multi-step excitations, are neglected. Such excitations would decrease the overall efficiency of the population transfer to the considered singlet-triplet mixture of excited states.}

The femtosecond impulse parameters are optimized to
obtain maximum yield in electronic states correlating with the first excited
atomic asymptote. We present a map of the optimization space
of the impulse parameters and the detailed population statistics.
Inspired by~\cite{Mizuno2018} we propose a modified
power law to describe the population decay. 

\section{\label{sec:level2} Computation method}

% {\textit{Computation method}} ---
The time propagation of a system of coupled time-dependent nuclear Schrödinger
equations (TDSE) for multiple electronic states and time-dependent Hamiltonian
follows the \textit{semi-global} method~\cite{Schaefer2017}. The coupling
between the electronic states caused by the electric field is time-dependent.
The TDSE has the following form

\newcommand*{\Hbf}{\hat{\mathbf{H}}} \newcommand*{\Vbf}{\hat{\mathbf{V}}}
\begin{equation}
  \label{eqSchTDHCoupled}
  i\hbar \frac{\partial}{\partial t}
  \begin{pmatrix}
    {\psi}_1 \\[1mm]
    {\psi}_2 \\[1mm]
    {\psi}_3
  \end{pmatrix}
  =
  \begin{bmatrix}
    \Hbf_1        & 0             & \Vbf_{1,3}    \\[1mm]
    0             & \Hbf_2        & \Vbf_{2,3}    \\[1mm]
    \Vbf_{3,1}    & \Vbf_{3,2}    & \Hbf_{3}   
  \end{bmatrix}
  \begin{pmatrix}
    {\psi}_1 \\[1mm]
    {\psi}_2 \\[1mm]
    {\psi}_3
  \end{pmatrix},
\end{equation}

\noindent
where the $\Hbf_{m}$ corresponds to the Hamiltonian for the respective
electronic state in the Born-Oppenheimer approximation

\begin{equation}
\Hbf_{m}=-\frac{\hbar^2}{2M}\nabla^2 + V_m(R)\hspace{10mm}m=1, 2, 3
\end{equation}

\noindent
and $\Vbf_{m,n}(t)$ are
either (1) the respective transition dipole moment functions $\mu$
multiplied\footnote{Following Tannor~\cite{Tannor2007} (page 396) we neglect the
vector character of $\mu$ \revB{because the non-rotating molecule (J=0)} is oriented along the \revB{propagation} direction of the \revB{linearly polarised laser} field.} by the impulse electric field $E(t)$ or (2) the spin-orbit potential $\xi$; depending on the nature of the coupling between the two electronic states as follows

\begin{equation}
\begin{split}
\Vbf_{1,3}(R,t)           &= \Vbf_{3,1}(R,t)           = -\mu_{1^3\Sigma^{+}-1^3\Pi}(R)\, E(t) \\
\Vbf_{2,3}(R)\phantom{,t} &= \Vbf_{3,2}(R)\phantom{,t} = \xi_{2^1\Sigma^+-1^3\Pi}(R).
\end{split}
\end{equation}

The Hamiltonian for three electronic states is summarized in
Tab.~\ref{tab:table1}.

\newcommand{\Cross}{$\circ$} \newcommand{\OK}{$+$}
\begin{table}[t]
  \caption{The schematic view of the Hamiltonian matrix of the system used in
    the calculations. The TDMFs between electronic states $\mu$ multiplied by the
    electric field $E(t)$ are marked with '$+$'. The lack of coupling between
    states is marked with '$\circ$'. The spin-orbit coupling is marked with SOC. The
    PECs are on the matrix diagonal.}
  \label{tab:table1}
  \begin{center}
    \begin{tabular}{l c c c}
      \toprule%  &                   &                     &            \\
      ~          & $1^{3}\Sigma^{+}$ & $2^{1}\Sigma^{+}$ & $1^{3}\Pi$ \\
      \midrule
      $1^{3}\Sigma^{+}$ & PEC               & \Cross              & \OK        \\
	    $2^{1}\Sigma^{+}$ & \Cross            & PEC                 & SOC        \\
	    $1^{3}\Pi$        & \OK               & SOC                 & PEC        \\
      \bottomrule
    \end{tabular}
  \end{center}
\end{table}

\begin{figure}[t]
  \includegraphics[width=\columnwidth,trim={12mm 22mm 25mm
    12mm},clip]{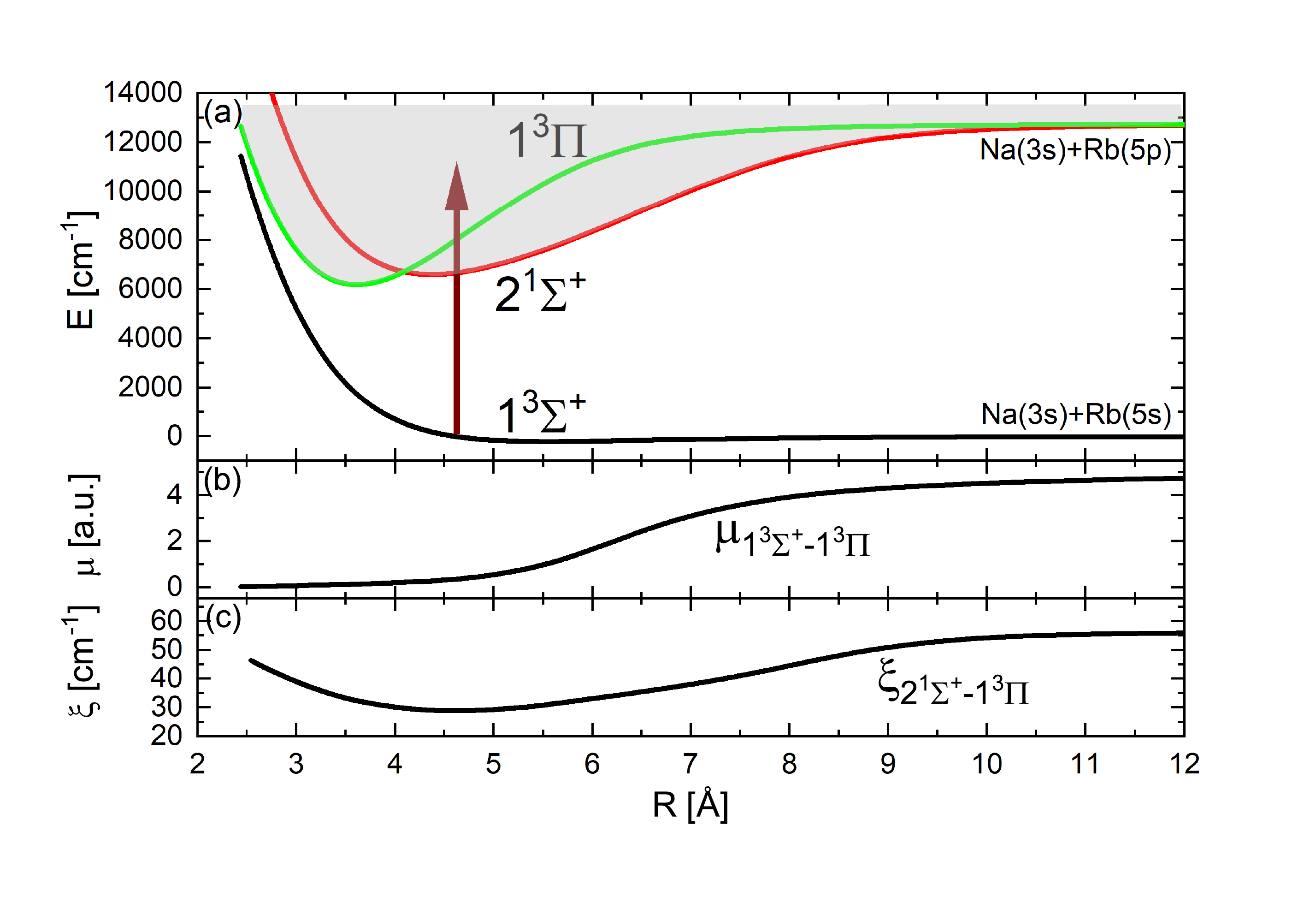}
  \caption{\label{fig:PEC} (a) electronic states of the NaRb molecule with a schematic arrow showing the transition imposed by the impulse and a gray area representing excited vibrational states; (b) the
    transition dipole moment function $\mu$ and (c) the spin-orbit coupling $\xi$.
    See Tab.~\ref{tab:table1} for the schematic view of the Hamiltonian in which
    they are used.}%
\end{figure}

The gist of the \textit{semi-global} method~\cite{Schaefer2017,Kozicki2023} is that the
evolution operator is expanded, using $K$ terms, in the Krylov space into two
parts: (1) the time-dependent part of the Hamiltonian and (2) the Hamiltonian in
the middle of the timestep $t+\frac{\Delta t}{2}$, for which the time-dependent part
serves as a correction calculated in the $M$ interior sub-timesteps. The
calculation of solutions for $M$ interior sub-timesteps is iterated several
times (usually two to five) within a single global timestep $\Delta t$ until the
solution converges with the requested error tolerance $\epsilon$. Then the calculation
moves on to the next timestep.

The \textit{semi-global} method is implemented in C++, thus allowing
calculations faster than the original Matlab code~\cite{Schaefer2017} and
extended with the ability to handle an arbitrary number of electronic states (a
feature not present in~\cite{Schaefer2017}) such as in
Eq.~\ref{eqSchTDHCoupled}. It is then used with the following parameters (which
are discussed in detail in~\cite{Schaefer2017}): the timestep $\Delta t=1$~a.u., the
number of interior Chebyshev time points $M=5$ and the number of expansion terms
used for the computation of the function of a matrix $K=7$. The error tolerance
is set to $\epsilon=10^{-10}$. The discretization grid has 6144 points and the
absorbing boundary condition (the same as in~\cite{Schaefer2017}) is placed at
the distance $R_{max}=70$~$a_0$.

\begin{figure}[t]
  \includegraphics[width=\columnwidth]{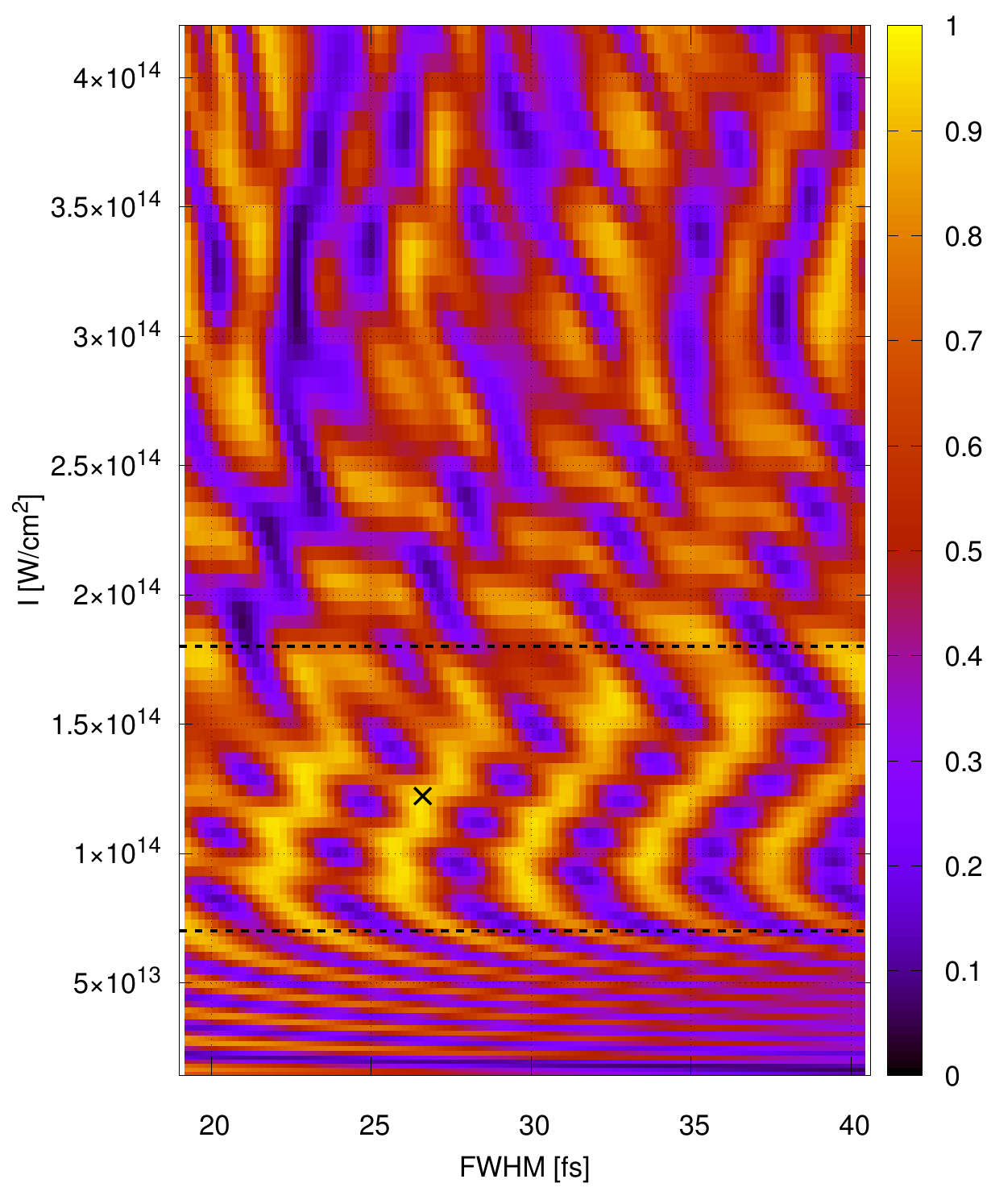}
  \caption{\label{fig:ScanHundA}Sum of the population on $2^{1}\Sigma^{+}$ and
    $1^{3}\Pi$ depending on the FWHM of the laser impulse and its intensity $I$.
    The peak population $P_0=97.2$\% at FWHM$=$26.6~fs,
    $I=1.2215\times10^{14}$~[W/cm$^2$] is marked with a cross. The horizontal bands
    separated by dashed lines are discussed in the text.}
\end{figure}

\section{\label{sec:level3} Results and discussion}

%{\textit{Results and discussion}} ---
Our previously calculated PECs~\cite{Wiatr2015}, TDMF $\mu_{1^3\Sigma^+-1^3\Pi}$ and SOC
$\xi_{2^1\Sigma^+-1^3\Pi}$ along with the schematic transitions are
shown in Fig.~\ref{fig:PEC}.

\begin{figure}[t]
\includegraphics[height=\columnwidth,angle=270]{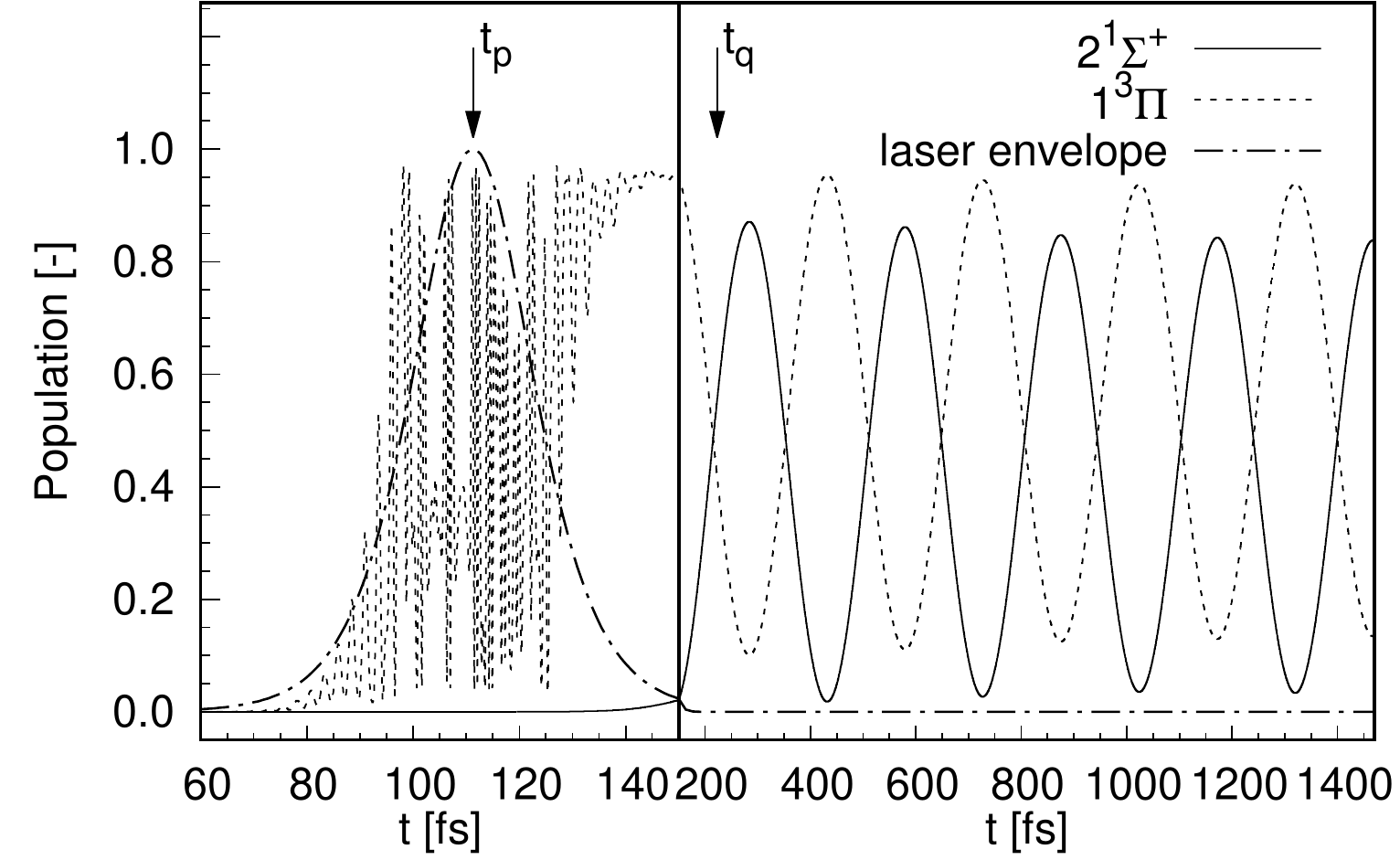}
\caption{\label{fig:Pop1} The evolution of the population on the two excited
  states $(2^1\Sigma^+$, $1^3\Pi)$. The impulse FWHM is 26.6~fs, the intensity is
  $1.2215\times10^{14}$~[W/cm$^2$] and the wavelength is 1560~nm. The impulse
  parameters correspond to the peak population found in the scan in
  Fig.~\ref{fig:ScanHundA}. The period of oscillation between the two levels is
  $296.35$~fs. The first peak of $2^1\Sigma^+$ is $172.49$~fs after $t_p$ (see
  Eq.~\ref{eq:laser}).
  \revA{The $t_q=222.53$~fs (9200~a.u.) marks the end of the laser impulse at which Fig.~\ref{fig:ScanHundA} is shown}.
  In the figure, the first 60~fs, when the impulse is weak
  and the population is near zero, are not shown for brevity.}
\end{figure}

As the initial condition, the $v=20$ eigenfunction of the $1^{3}\Sigma^{+}$
state is used, because it is one of the highest populated levels obtained in the
experiment~\cite{Guo2016,Guo2017}. Then the following unchirped impulse is used:

\begin{equation}
\label{eq:laser}
E(t)=E_o\,\, \textrm{sech}^2\left(({t-t_p})/{\tau_p}\right) \cos(\omega t),
\end{equation}

\noindent
where the frequency $\omega$ corresponds to a commercially available laser with the
wavelength $\lambda=1560$~nm, the impulse center $t_p=111.26$~fs (4600~a.u., also
marked in Fig.~\ref{fig:Pop1} with an arrow), while the maximum intensity $I=\varepsilon_0
c E_0^2/2$ and the full width at half maximum (FWHM) are optimized to maximize
the population on the excited state (the FWHM of the impulse equals to $1.76 \tau_p$).
\revA{The laser $\textrm{sech}^2$ envelope is shown in Fig.~\ref{fig:Pop1}}.
The optimization is performed by scanning a range of these two parameters. The
optimal FWHM was found to be equal to 26.6~fs and is comparable to the period of
vibration motion of the molecule. This requires potential energy curves that decribe electronic
states to be accurate in dynamics calculations. The laser intensity scan range
extends from 10$^{10}$ to 10$^{15}$ W/cm$^2$. In our three-states model, an
intensity of the order 10$^{14}$ gives the highest population in the upper
triplet state. This is the starting point for calculating the population
distribution. Despite the limitations of the model of the molecular structure,
they provide important physical information about the decay process.

The scan of the sum of the population on two excited states $2^{1}\Sigma^{+}$ and
$1^{3}\Pi$ is shown in Fig.~\ref{fig:ScanHundA}
\revA{at time $t_q$=222.53~fs (9200~a.u.) marked in Fig.~\ref{fig:Pop1} as the end of the laser impulse}.
We note the existence of three characteristic horizontal bands. The first band
occurs for the intensity $I<0.7\times10^{14}$~[W/cm$^2$]. Here the horizontal lines
of the higher population indicate that the FWHM energy spread of the impulse
does not affect the population and it mostly depends on the impulse intensity.
The second band is in the middle range of $I$ between $0.7\times10^{14}$ and
$1.8\times10^{14}$~[W/cm$^2$]. Here a semi-diagonal pattern emerges where the FWHM
energy spread dependence plays a major role. Finally in the third range for
$I>1.8\times10^{14}$~[W/cm$^2$] a chaotic pattern emerges.

Next, using the optimal parameters $I$ and FWHM (marked with a cross in
Fig.~\ref{fig:ScanHundA}) corresponding to the maximum population, the time
evolution of the population happens mainly on the excited electronic states.
Fig.~\ref{fig:Pop1} shows results for obtaining the maximum population found of the $1^3\Pi$ excited state, which is $P_0=97.2$\%.

The effect of the spin-orbit coupling is visible in Fig.~\ref{fig:Pop1} as the
population exchange between electronic states $2^1\Sigma^+$ and $1^3\Pi$. The period of
the oscillation between the two states is $296.35$~fs. The first maximum on
$2^1\Sigma^+$ occurs $172.49$~fs after the center of the impulse $t_p$. In our model,
this would be the optimal time delay for the next impulse to dump the molecule
from the $2^1\Sigma^+$ state to the absolute ground state using the dumping impulses occurring with a period of $296.35$~fs. It means that the peaks of the impulses should correspond to the peaks of the population in the $2^1\Sigma^+$ state.

\begin{figure}[t]
\includegraphics[height=\columnwidth,angle=270]{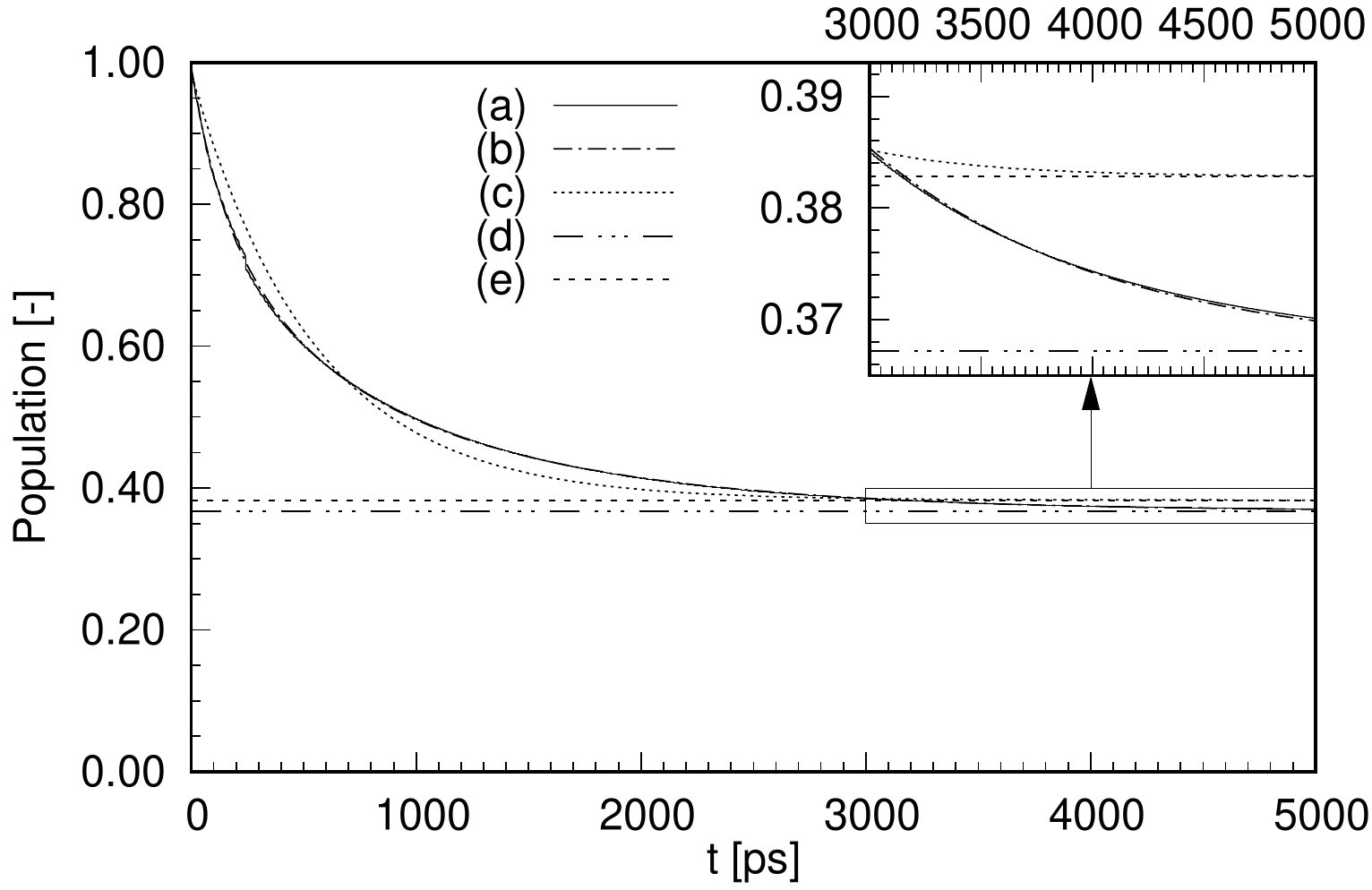}
    \caption{\label{fig:PowerLaw1}
    Sum of population on $2^{1}\Sigma^{+}$ and $1^{3}\Pi$ decaying during the first 5000~ps.
    The decay starts at $t_0=10$~ps.
    (a) numerical result;
    (b)~best fit using the power-law in Eq.~\ref{eq:PowerLaw1};
    (c) best fit using the exponential decay law in Eq.~\ref{eq:ExpLaw1};
    (d) The best fit value of the final population $P_{f,pow}$ in Eq.~\ref{eq:PowerLaw1};
    (e) The best fit value of the final population $P_{f,exp}$ in Eq.~\ref{eq:ExpLaw1}.}
\end{figure}

The population between the electronic states is constantly \revA{exchanged} due to the
spin-orbit coupling. Figure~{\ref{fig:PowerLaw1}} shows the decay of the
population within the first nanosecond while the occurrence of spontaneous
emission to the ground state is still negligible. The higher frequencies in the wavepacket corresponding to energies that are above the first excited atomic asymptote are undergoing a dissociation. This process is slowed down due to the evolution of the wavepacket on the coupled potential of two excited electronic states. This causes a
population decay over a larger timescale and only a fraction of the original
population remains at the end, as shown in Fig.~{\ref{fig:PowerLaw1}}\footnote{\revA{Fig.~{\ref{fig:PowerLaw1}} took 6 months to calculate on single thread on processor AMD} \revA{EPYC 7702P with 3.35 GHz boost clock speed, with discretization grid having 6144 points per electronic level.}}.
As we show below, this is not an exponential decay as in the predissociation
process~\cite{Jasik2018a}. Instead, it is a distribution of the power law,
as it happens by the coincidence of population exchange between two
states~\cite{Mizuno2018,Moller2000,Shapiro1998,Brems2002}.

Following the idea in~\cite{Mizuno2018}, that the population decay in the case
of two coupled states should follow a power-law decay, we analyze the population
decay on the two excited states $2^{1}\Sigma^{+}$ and $1^{3}\Pi$ in our model. The
calculated population from Fig.~\ref{fig:PowerLaw1} is fitted respectively to
the modified power-law decay formula:

\begin{equation}
\label{eq:PowerLaw1}
P(t)=P_0 \left(\left(1-P_{f,pow} \right)\left({\frac{\tau_{pow}}{t+\tau_{pow}-t_0}}\right)^{\alpha t + \beta} + P_{f,pow} \right)
\end{equation}

\noindent
and to the exponential decay formula:

\begin{equation}
\label{eq:ExpLaw1}
P(t)=P_0 \left(\left(1-P_{f,exp} \right)\exp\left(-\frac{t-t_0}{\tau_{exp}}\right) + P_{f,exp} \right),
\end{equation}

\noindent
where $P_0=97.2$\% is the initial population on both states, $t_0=10$~ps is the
start of the decay (when the wavepacket reaches the absorbing potential),
$\tau_{pow}$, $\tau_{exp}$ are the fitting decay parameters, and $P_{f,pow}$ and
$P_{f,exp}$ are the fitting parameters describing the final non-zero population.
Additional parameters, $\alpha$ and $\beta$, are added to improve the fit, whereas
in~{\cite{Mizuno2018}} used a square root in this place. The $(\tau_{pow}/\bullet)$ part
of Eq.~\ref{eq:PowerLaw1} was modified with respect to~\cite{Mizuno2018} with an
extra $\tau_{pow}$ in the denominator in order to shift the function to the left so
that the initial population is not infinite when $t=t_0$ (as is the case
in~\cite{Mizuno2018}), but instead equals $P_0$. Also in both equations for
decay we used a modification of type: $( 1 - P_{f})\bullet + P_{f}$ to allow extra
fitting parameters $P_{f,pow}$ and $P_{f,exp}$ for the final non-zero
population. It shall be noted that compared to~\cite{Mizuno2018} the situation
here is also different: both potentials $2^{1}\Sigma^{+}$ and $1^{3}\Pi$ are bonding
potentials and neither of them is dissociative (see Fig.~\ref{fig:PEC}).
Therefore, we use modified power-law decay to fit our numerical results.

The fit results in Fig.~\ref{fig:PowerLaw1} are as follows:
$\tau_{pow}=215.86\pm0.17$~ps, $P_{f,pow}=0.3776\pm10^{-5}$ (the value on
Fig.~\ref{fig:PowerLaw1} is $0.3776\times97.2\%=36.7\%$),
$\alpha=2.004\times10^{-4}\pm10^{-7}$, $\beta=0.6979\pm0.0004$ and $\tau_{exp}=543.0\pm0.3$~ps,
$P_{f,exp}=0.3937\pm0.0001$ (the value on Fig.~\ref{fig:PowerLaw1} is
$0.3937\times97.2\%=38.2\%$).
For the power law RMSE$=0.0067$ and $R^2=0.996$, while for the exponential fit RMSE$=0.0185$ and $R^2=0.972$.
The best fit for exponential law produces a nonphysical
result because the fitted final population $P_{f,exp}$
(Fig.~\ref{fig:PowerLaw1}e) has value greater than the steadily decreasing
population (a). The fit for the power-law (Fig.~\ref{fig:PowerLaw1}b) is
significantly better, since it almost overlaps the numerical result from the
last timestep (Fig.~{\ref{fig:PowerLaw1}}a). The final population from the
modified power law (Fig.~\ref{fig:PowerLaw1}d) is $P_{f,pow}\times97.2\%=36.7\%$.
At 5~ns the calculated population dropped down to 37\%. Overall, the modified
power law provides a better description of the decay occurring in the spin-orbit-mediated dissociation process.

\section{\label{sec:level4} Conclusions}

%{\textit{Conclusions}} ---
In our study, we show the time evolution on the three coupled potential energy
curves of the diatomic alkali metals system -- NaRb. The necessary potential energy curves were taken from our earlier ab initio calculations and are presently supplemented with electronic transition dipole moment and spin-orbit coupling between the $2^{1}\Sigma^{+}$ and $1^{3}\Pi$ states. The dynamics of the considered system is driven by a
femtosecond impulse. The two parameters describing the impulse, i.e. intensity
and FWHM, are independently optimized, in order to obtain the largest
possible population of molecular states correlated with the first excited atomic
asymptote. We also identify the period of population oscillation between the two
excited states $2^1\Sigma^+$ and $1^3\Pi$.

This allows us to estimate the optimal time delay between the pump and dump
impulses, in order to show the possibility of obtaining the NaRb molecule in the singlet ground electronic
state. In this approach, the optimal time delay after a pump pulse to dump the
molecule to the ground state is 172.49~fs using a pulse with a period of 296.35~fs. We are also examining the combined population decay from the two excited
singlet and triplet states. We fit the decay of the population to a modified
power law and show that this fit describes this process better than an
exponential decay. We emphasize that these results can provide insight into
the quantum dynamical processes in diatomic molecules, where time plays a key
role. Our approach is possible to apply to even more complex systems where one
soft bond (e.g.~between two fragments in a molecule) is
most important to describe a time-dependent process. We also confirm that impulse control can be applied to the spin-orbit coupled states.

The possible extension of the calculations presented here is to replace the
femtosecond excitation with a femtosecond laser impulse with intensity and FWHM
commonly used in experiments as well as taking into account the permanent dipole
moment and light polarisation. Furthermore, an optimisation of the second
femtosecond laser impulse to obtain an ultracold NaRb dimer can be performed.
Finally, it would be possible to include higher electric and magnetic transition
moments, in order to study extremely high intensities.

All dynamic results are obtained from our new computer code which uses a
\textit{semi-global} method to expand the time evolution
operator. The method is improved to work with multiple electronic states and
time-dependent couplings. It can be a valuable tool for studying quantum
dynamics and planning future experiments.

\section*{Author Information}
\noindent{\textbf{Corresponding Author}}\\
\noindent * E-mail: \href{mailto:patryk.jasik@pg.edu.pl}{patryk.jasik@pg.edu.pl}\\[1mm]

\noindent{\textbf{ORCID}}\\
\newcommand{\OrcidLink}[1]{\href{https://orcid.org/#1}{#1}}
\noindent Janek Kozicki:        \OrcidLink{0000-0002-8427-7263}\\
\noindent Patryk Jasik:         \OrcidLink{0000-0002-6601-0506}\\
\noindent Tymon Kilich:         \OrcidLink{0000-0001-6831-694X}\\
\noindent Józef E. Sienkiewicz: \OrcidLink{0000-0002-1149-3846}\\[1mm]

\noindent{\textbf{Conflicts of interest}}\\
%The authors declare no competing financial interest.
There are no conflicts of interest to declare

\section*{Data availability}
\noindent The data that support the findings of this study are publicly available in the Bridge of Knowledge Open Data Repository,\\ \url{https://doi.org/10.34808/kkhq-s579}.

\section*{Acknowledgements}
We acknowledge the partial support by the Polish Ministry of Education and Science, COST action CA18222 ``Attosecond Chemistry'', and COST action CA21101 ``Confined Molecular Systems: From a New Generation of Materials to the Stars''. Calculations have been carried out using resources provided by the Centre of Informatics Tricity Academic Supercomputer \& Network and Wroclaw Centre for Networking and Supercomputing.

\bibliography{manuscript.bib}% Produces the bibliography via BibTeX.

\end{document}